# Plasmonic Electro-Optic Modulators based on Epsilon-Near-Zero Materials: Comparing the Classical Drift-Diffusion and Schrödinger-Poisson Coupling Models

MASOUD SHABANINEZHAD,[1,2,3] HAMID MEHRVAR,[4] ERIC BERNIER,[4] LORA RAMUNNO,[2,3] AND PIERRE BERINI[1,2,3]

[1]School of Electrical Eng. and Comp. Sci., University of Ottawa, Ottawa, ON K1N 6N5, Canada
[2]Department of Physics, University of Ottawa, Ottawa, ON K1N 6N5, Canada
[3]NEXQT Institute, 25 Templeton St., Ottawa ON, K1N 6N5, Canada
[4]Huawei Technologies Canada, Canada, Kanata, ON K2K 3J1, Canada

**Abstract:** We present the design, modeling, and optimization of high-performance plasmonic electro-optic modulators leveraging voltage-gated carrier density in indium tin oxide (ITO) where the gated carrier density is modeled using both the Classical Drift-Diffusion (CDD) and Schrödinger-Poisson Coupling (SPC) methods. The latter ensures a more detailed and precise description of carrier distributions under various gate voltages which gains particular significance when applied to an epsilon-near-zero (ENZ) medium such as ITO. Combining the nanoscale confinement and field enhancement enabled by surface plasmon polaritons with the ENZ effect in ITO, modulator designs integrated with silicon waveguides and optimized for operation at $\lambda_0 = 1550$ nm achieve a 3-dB bandwidth of 210 GHz, an insertion loss of 3 dB, and an extinction ratio of 5 dB for an overall length of < 4 µm as predicted by the SPC model. Our results illustrate trade-offs between high-speed modulator operation and low insertion loss, *vs.* extinction ratio, and the need for precise modelling of carrier distributions in ENZ materials.

## 1. Introduction

The development of high-speed, power efficient, and compact electro-optic modulators is critical for advancing integrated photonic circuits, which are essential for next-generation communication systems [1–4]. Traditional semiconductor-based modulators, such as those made from silicon, germanium, and compound semiconductors, face significant challenges [5–8]. Silicon-based modulators suffer from large device footprints, narrow intrinsic bandwidth and limited temperature tolerances [5,6], while germanium and compound semiconductor modulators encounter integration issues with existing silicon photonics platforms [7,8]. These limitations hinder their effectiveness in applications requiring high modulation speeds and broad optical bandwidths integrated on-chip [9]. Recent advancements in plasmonic modulators have shown promise in overcoming these limitations [1–4,10–12]. Plasmonic modulators leverage the unique properties of surface plasmon polaritons (SPPs) to achieve subwavelength confinement and enhanced light-matter interaction[1–4,10–12]. This enables faster response times, smaller footprints, and potentially, compatibility with complementary metal-oxide-semiconductor (CMOS) processes. Studies have highlighted the potential of plasmonic modulators to achieve high modulation speeds, making them suitable for high-throughput optical interconnects and other demanding broadband applications.

Several approaches have been explored in plasmonic modulators [3,9,11–15]. Studies generally focus on optimizing the device architecture, including the use of novel waveguide configurations and advanced fabrication techniques, to improve modulation efficiency and bandwidth. For instance, using an electro-optic polymer, Hoessbacher *et al.* demonstrated a plasmonic modulator with an active length of 12.5 μm, a bandwidth exceeding 170 GHz, insertion losses around 8 dB, and an extinction ratio of 4.7 dB [3]. Other researchers have investigated different material systems, such as graphene and transition metal dichalcogenides, to exploit their tunable optical properties and high carrier mobility [9,13,14].

One of the key mechanisms leveraged in plasmonic modulators is the epsilon-near-zero (ENZ) effect in transparent conductive oxides, like indium tin oxide (ITO) [12,16–20]. The ENZ effect occurs when the real part of the permittivity approaches zero, resulting in unique optical properties such as enhanced electro-optic and nonlinear interactions, and strong field confinement [21,22]. This effect can be exploited by incorporating ITO in a MOS (metal-oxide-semiconductor) structure to achieve modulation of the modal effective refractive index with applied voltage [12,16–20].

Previous work on plasmonic modulators exploiting the ENZ effect has relied solely on the Classical Drift Diffusion (CDD) model to simulate the carrier density within the perturbed region of ITO. We recently emphasized the need to employ the Schrödinger-Poisson Coupling (SPC) model [22], which incorporates Schrödinger's equation into the CDD model in a self-consistent manner [23,24], when modeling carrier distributions in an ENZ material, as the latter captures effects due to carrier quantization and is fundamentally more accurate.

In this paper we apply the CDD and SPC models to the design of plasmonic modulators based on an ENZ material (ITO) and emphasize the differences in the predictions made by both models. The carrier distribution under bias obtained by the SPC model yields two ENZ crossings within the perturbed region compared to only one for the CDD model – this qualitative difference leads to significant quantitative differences in the predicted performance of the modulators. We find that using the CDD model underestimates perturbations in the ENZ region relative to the SPC model, with the latter leading to significantly improved device designs. Although applied to modulators, our conclusions should extend to other electro-optic devices exploiting ENZ effects, plasmonic or otherwise.

## 2. Structure of interest

The proposed modulator structure consists of two active back-to-back mode conversion tapers of length $L_T$, each composed of a pair of metal-insulator-metal (MIM) stacks strategically positioned on top of a planarized Si waveguide (violet), as depicted in Fig. 1(a). The modulator operating free-space wavelength is selected as $\lambda_0$ = 1550 nm. The bottom metal of the MIM stacks is selected as a 30 nm thick layer of Al (blue), upon which a 3.6 nm thick layer of alumina ($Al_2O_3$, orange) is assumed grown. A 20 nm thick layer of ITO layer (green) topped by a 50 nm thick Au layer (gold) completes the stacks. The bottom Al layer acts as the anode, whereas the top Au layer forms the cathode, creating a simple and effective electrical driving scheme. Thus, the MIM stacks operate as metal oxide semiconductor (MOS) capacitors formed by the bottom metal layer (Al), the oxide layer ($Al_2O_3$), and the ITO layer acting as the semiconductor. Although Al exhibits higher optical absorption than Au, it provides significant fabrication advantages. Specifically, a high quality $Al_2O_3$ layer can be easily created on the surface of Al by thermal oxidation [25,26], avoiding the challenges with deposition, such as inconsistent film quality and island formation [27]. The low relative permittivity of $Al_2O_3$ (9.3 [28]) compared to, *e.g.*, $HfO_2$ (25 [29]), ensures that the capacitance per unit length is manageable.

Fig. 1(b) shows a detailed cross-sectional view of the modulator over the area bounded by the black dotted rectangle in Fig. 1(a), identifying each layer's dimensions and material for improved clarity. The material stack is assumed fabricated using standard wafer fabrication techniques, with a gap of length g between the MIM sections, defined using focused ion beam (FIB) milling. To accommodate for fabrication deviations, we incorporated an over-milling depth into the Si of $t_{mill} = 20$ nm.

The modulation process commences by launching the fundamental transverse magnetic $TM_0$ mode of the Si waveguide toward the modulator. This mode is then adiabatically transferred into the MIM sections via an input taper, during which the gap between the MIM stacks, linearly narrows from 520 nm to a minimum value ($g_{min}$) in the middle of modulator. The light is then returned adiabatically to the $TM_0$ mode of the Si waveguide through an output taper, where the gap linearly widens back to 520 nm. This linear adjustment of the gap ensures efficient mode conversion and minimizes losses during signal transmission, optimizing the device's overall performance. Driving these back-to-back tapers into accumulation increases the attenuation of the MIM mode, and thus the insertion loss of the tapers, thereby modulating the intensity of the transmitted light.

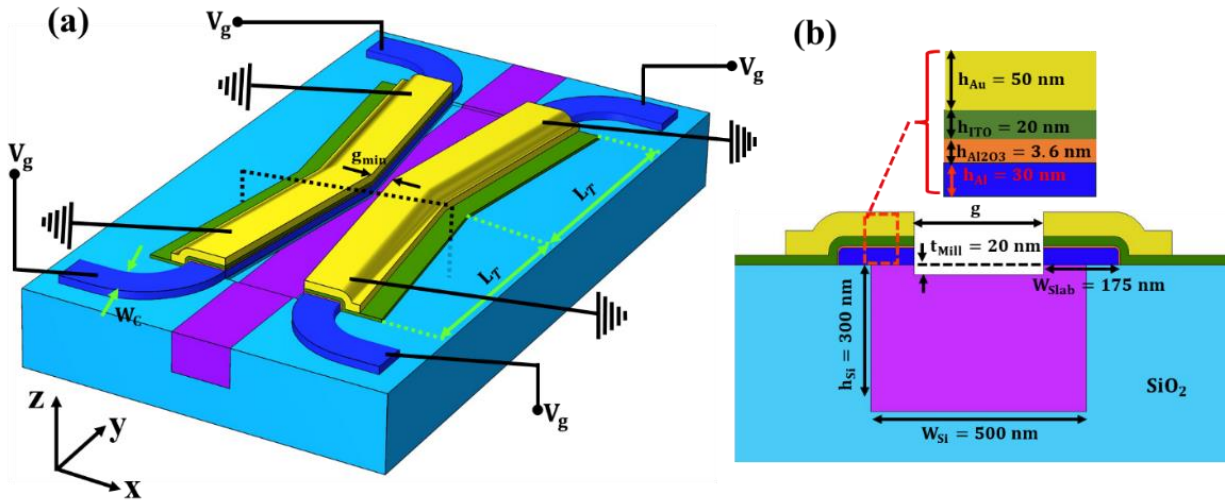

**Fig. 1.** (a) 3D schematic of the plasmonic modulator, featuring active back-to-back mode conversion tapers on a planarized Si (violet) waveguide, and four voltage connection points. (b) Cross-sectional view detailing the layer dimensions of the device, over the area bounded by the black dotted rectangle in Fig. 1(a).

## 3. Theoretical Modeling

### 3.1 Electrostatic Modeling

The core architecture of our high-speed plasmonic modulator consists of an $Al - Al_2O_3 - ITO - Au$ stack, forming a MOS structure within the MIM configuration. Applying voltage across this MOS structure perturbs the carrier density in the ITO layer, inducing either depletion or accumulation depending on the voltage polarity. These modifications significantly impact the optical properties of the modulator by dynamically altering the complex refractive index of ITO. To simulate the perturbed carrier density within ITO, we utilized two models: the Classical Drift-Diffusion (CDD) and the Schrödinger-Poisson Coupling (SPC) models [22]. The latter ensures a more detailed and precise depiction of electro-optical interactions within the device compared to the former. Given that the perturbed carrier density varies with distance from the oxide-ITO interface, a 1D model is sufficient for accurate predictions of the perturbed carrier density, avoiding the computational burden associated with higher-dimensional models.

The CDD model is utilized to simulate the macroscopic carrier distribution within the ITO layer, effectively capturing the transport phenomena under the influence of an applied voltage. This model incorporates the current density, continuity and Poisson's equations, which describe how carriers move and interact within the semiconductor and links the electric potential distribution to the charge distribution within the material [12,22,30]. The electron and hole concentrations within the ITO are governed by the following relationships:

$$N(d_{ox}) = N_c F_{1/2}\left(\frac{E_{fN}-E_c}{k_B T/q}\right) \tag{1a}$$

$$P(d_{ox}) = N_v F_{1/2}\left(\frac{E_v-E_{fp}}{k_B T/q}\right) \tag{1b}$$

Here, $d_{ox}$ represents the distance from $Al_2O_3 - ITO$ interface into ITO; $N_c$ and $N_v$ are the effective density of states for the conduction and valence bands, respectively; $E_{fN}$ and $E_{fP}$ are the quasi-Fermi levels for electrons and holes; and $E_c$ and $E_v$ are the edges of the conduction and valance bands. $k_B$ is Boltzmann's constant, T is the temperature, q is the elementary charge, and $F_{1/2}$ is the Fermi-Dirac integral [31].

However, at length scales comparable to the electron wavelength, quantization effects such as confinement, compressibility, and tunneling significantly influence device characteristics [22–24]. To accurately capture these phenomena along with electron diffraction and interference effects - which classical models like the CDD model do not address - we have adopted the SPC model [22–24].

The SPC model operates by iteratively solving the Schrödinger and Poisson equations to achieve a self-consistent solution [22–24]. The process begins with an initial potential distribution and carrier density profile derived from the CDD model. The electrostatic potential ϕ, fulfilling the nonlinear Poisson equation, is coupled to the Schrödinger equation through the potential energy term $V_e = q\phi$. This results in the one-dimensional Schrödinger equation for a MOS structure expressed as:

$$-\hbar^2 \nabla \cdot \left(\frac{\nabla \Psi_j(d_{ox})}{2m_{eff}}\right) + (V + V_e)\Psi_j(d_{ox}) = E_j \Psi_j(d_{ox}) \tag{2}$$

Here, $m_{eff}$ denotes the effective mass of the electron, V is the confinement potential, and $\Psi_j$ is the $j^{th}$ normalized eigenfunction corresponding to the eigenenergy $E_j$. The carrier density profile N is then computed using a statistically weighted sum of the probability densities:

$$N(d_{ox}) = \sum_j W_j |\Psi_j(d_{ox})|^2 \quad (3)$$

where $W_j$ are weight factors calculated for one-dimensional structures [22]. Following the calculation of $N(d_{ox})$ from Eq. 3, we compute the space charge density ρ. The calculated ρ is then utilized to update the potential distribution ϕ by solving Poisson's equation. This iterative process is continued, refining ϕ until a converged solution is reached.

To simulate the perturbed carrier density within ITO at various voltages, we utilized simulation features within COMSOL MULTIPHYSICS 6.1 [32]. For the CDD model, we used the Semiconductor Physics interface from the Semiconductor Module along with the Semiconductor Equilibrium study. For the SPC model, we employed the Schrödinger Equation and Electrostatic Physics interfaces under the Semiconductor Module. These models were integrated using the Schrödinger-Poisson Coupling physics feature. To run the SPC simulations, we utilized two study types: the Stationary study, which handles static simulations, and the Schrödinger-Poisson Coupling study, which captures the dynamic interaction between the electrostatic potential and quantization behavior of carriers. Detailed guidance on these methods, setup, and boundary conditions are provided in [22].

In our simulations, the thickness and the work function of the Al were set to 30 nm and 4.1 eV [33], respectively. The Al$_2$O$_3$ layer was specified with a thickness of 3.6 nm and static relative permittivity of 9.3 [28]. For the ITO layer, the thickness was set to 20 nm, the bulk doping concentration ($N_b$) was taken as $2.65 \times 10^{20}$ (cm$^{-3}$) [22], the bandgap energy ($E_g$) was set to 2.8 eV [29], the effective mass of the electrons was 0.35 $m_e$ [29] where the $m_e$ is the free electron mass, and the electron affinity ($\chi_s$) and static relative permittivity ($\varepsilon_s$) were taken as 4.8 eV and 9.1, respectively [29]. Since the Fermi energy of Al is higher than ITO, short-circuiting the terminals results in electron flow from Al to ITO to align the Fermi levels, resulting in the creation of an accumulation region within ITO at zero applied voltage. The structure reaches the classical flat band at a gate voltage of -0.26 V. The maximum gate voltage that can be applied is restricted by the breakdown field ($E_{bk}$) of Al$_2$O$_3$, which is taken as $E_{bk} = 10$ MV/cm [28]. Therefore, we limit the gate voltage to the range of −3 V to 3 V to ensure the electric field in Al$_2$O$_3$ stays below the breakdown threshold.

*3.2 Optical Modeling*

The carrier density distributions obtained from the CDD and SPC models at various applied voltages, $N(d_{ox})$, were used in the Drude model to compute the corresponding permittivity distributions in ITO at $\lambda_0 = 1550$ nm:

$$\varepsilon(d_{ox}, \omega) = \varepsilon_\infty - \left(\frac{N(d_{ox})\omega_{b,p}^2}{N_b}\right) / (\omega^2 + i\Gamma\omega) \quad (4)$$

where, $\varepsilon_\infty = 3.92$ is the high-frequency relative permittivity, $N_b = 2.65 \times 10^{20}$ cm$^{-3}$ is the unperturbed carrier density, $\omega_{b,p} = 1.55 \times 10^{15}$ rad/s denotes the bulk plasma frequency, and $\Gamma =$ $4.4 \times 10^{13}$ rad/s is the damping frequency. These parameters were derived by fitting the measured bulk permittivity of ITO to Eq. 4 [22].

To understand the modal transformation of the modulator under varying drive voltage and design conditions, the frequency-domain vector wave equations, derived from Maxwell's equations, were solved subject to specific boundary conditions [34]:

$$\nabla \times \left(\frac{1}{\mu_r} \nabla \times \boldsymbol{E}(r)\right) - k_0^2 \varepsilon_r(r) \ \boldsymbol{E}(r) = 0 \tag{5}$$

$$\nabla \times \left(\frac{1}{\varepsilon_r(r)} \nabla \times \boldsymbol{H}(r)\right) - k_0^2 \mu_r \boldsymbol{H}(r) = 0 \tag{6}$$

Here, $\mu_r = 1$ is the relative permeability, and **E** and **H** represents the electric and magnetic field vectors, respectively. $\varepsilon_r(r)$ is the voltage-modulated spatially dependent relative permittivity (Eq. (4)), and $k_0$ is the free-space wavenumber.

For the optical simulations, the Waveoptics module in COMSOL 6.1 was employed [32]. To compute the modes in the modulator cross-section depicted in Fig. 1(b) under various design and operating conditions, the 2D Electromagnetic Waves (frequency domain) interface was used. The refractive index of the materials at $\lambda_0$ = 1550 nm were taken as follows: $n_{Au} = 0.238 + 11.26i$ [35], $n_{Al} = 1.347 + 14.13i$ [35], $n_{Al_2O_3} = 1.75 + 0.0i$ [36], $n_{Si} = 3.44 + 0.0i$ [37], $n_{SiO_2} = 1.44 + 0.0i$ [38] and $n_{Air} = 1.0 + 0.0i$.

For comprehensive 3D simulations of the entire modulator shown in Fig. 1(a), and to analyzing adiabatic mode conversion in the tapers at different voltages, the Beam Envelope method in COMSOL 6.1 was employed [32]. This method models complex electromagnetic fields over large spatial domains without requiring the fine spatial resolution of full-wave simulations. It approximates the electromagnetic field as a slowly varying envelope, multiplying a rapidly oscillating carrier wave, thus reducing computational load while preserving simulation accuracy.

In the Beam Envelope method, the electric and magnetic fields are expressed as:

$$\boldsymbol{E}(r) = \boldsymbol{E}_{env}(r)e^{-i(\phi(r))} \tag{7}$$

$$\boldsymbol{H}(r) = \boldsymbol{H}_{env}(r)e^{-i(\phi(r))} \tag{8}$$

where $\boldsymbol{E}_{env}(r)$ and $\boldsymbol{H}_{env}(r)$ are the slowly varying electric and magnetic field envelopes, respectively, and $\phi(r)$ is the phase function of the mode. By substituting these equations into Maxwell's equations, the simplified wave equations for the Beam Envelope are obtained [32]:

$$(\nabla - i\nabla\phi(r)) \times \left[\frac{1}{\mu_r}(\nabla - i\nabla\phi(r)) \times \boldsymbol{E}_{env}(r)\right] - k_0^2 \varepsilon_r(r) \ \boldsymbol{E}_{env}(r) = 0 \tag{9}$$

$$(\nabla - i\nabla\phi(r)) \times \left[\frac{1}{\varepsilon_r(r)}(\nabla - i\nabla\phi(r)) \times \boldsymbol{H}_{env}(r)\right] - k_0^2 \mu_r \ \boldsymbol{H}_{env}(r) = 0 \tag{10}$$

To ensure the integrity of our simulation results, the outer boundaries terminating our simulation cross-sections and volume were placed sufficiently far to prevent any influence on the computations. Scattering boundary conditions were implemented at these boundaries in our 2D and 3D simulations within COMSOL to effectively model outgoing waves. Numerical ports were integrated at the input and output of the structure in the 3D simulations to excite the TM$_0$ mode and measure the transmittance, respectively. To mitigate back reflections at the input and output ports and to ensure minimal absorption of the transmitted light, the boundaries of these ports were extended. Perfectly Matched Layers (PMLs) were subsequently applied at the ends of these extended regions. By distancing the PMLs from the modulator, we effectively minimize the interference between the PMLs and the ports.

## 4. Results and Discussion

### 4.1 Modal Transformation

We conducted modal analyses on the modulator cross-section depicted in Fig. 1(b). This analysis focused specifically on modal evolution of the $TM_0$ mode of the Si waveguide. The analysis was initiated at a gap width of g = 520 nm and tracked the evolution of this mode as the gap was gradually reduced to 340 nm. The modulator was assumed biased at the flat band voltage ($V_{fb} = -0.26$ V).

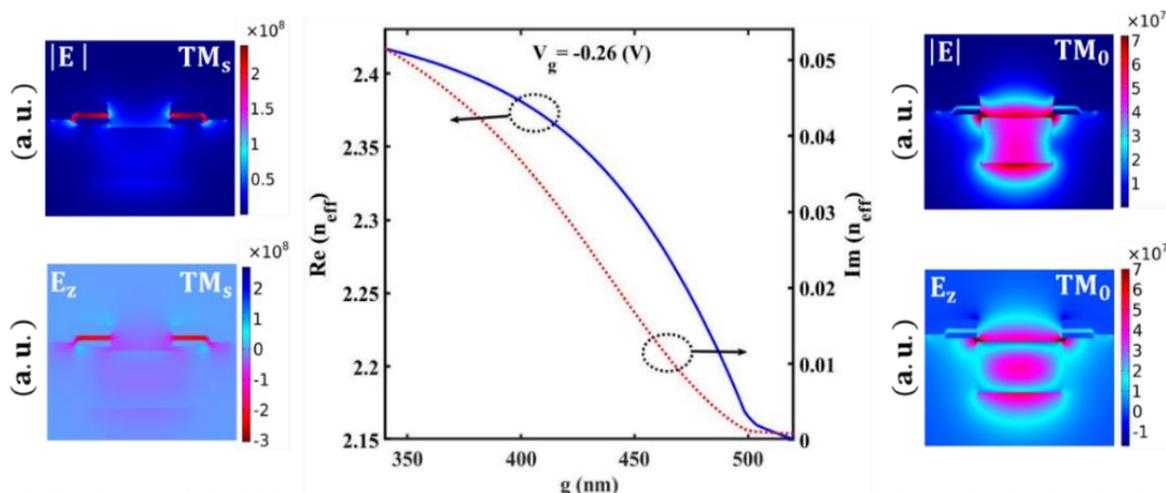

**Fig. 2.** Evolution of the $TM_0$ mode in the Si waveguide into the MIM coupled mode at the flat band voltage ($V_{fb} = -0.26$ V). The main panel plots changes in the real and imaginary parts of the effective refractive index as the gap decreases from 520 to 340 nm, with corresponding electric field profiles (presented in arbitrary units (a.u.)) at the end points.

Fig. 2 plots the evolution of the real and imaginary parts of the effective refractive index of this mode as the gap decreases from 520 to 340 nm. As Fig. 2 reveals, significant increases in the real and imaginary parts of the refractive index occur as the gap narrows below the width of the Si waveguide (g < 500 nm). The real part of the effective index ($Re(n_{eff})$) increases from 2.15 to 2.42, indicating a substantial increase in optical confinement as the gap narrows. Simultaneously, the imaginary part of the effective index ($Im(n_{eff})$), which characterizes the attenuation within the MIM structure, exhibits a significant increase from $8.10 \times 10^{-4}$ to $5.15 \times 10^{-2}$. This dramatic rise in attenuation indicates increased interaction between the light and the MIM stacks as the gap diminishes.

The electric field profile and the distribution of the main transverse electric field component of the $TM_0$ mode are plotted on the right side of Fig. 2 for g = 520 nm, whereas the field profile and the distribution of the main component of the MIM mode are shown on the left side of the figure for g = 340 nm. All fields are presented in arbitrary units (a.u.). The $TM_0$ mode of the Si waveguide, observed at g = 520 nm, transitions smoothly into the symmetric coupled TM mode supported by the MIM waveguides at $g = 340$ nm, where the main component of the electric field, $E_z$, is symmetrically distributed over the cross-section of the MIM stacks. The fact that the $TM_0$ mode of the Si waveguide evolves smoothly into the symmetric coupled TM mode of the MIM waveguides implies that adiabatic tapers can be designed.

*4.2 Taper Optimization at Flat band Voltage*

The taper sections are crucial for the efficient and selective transformation of modes between the Si waveguide and the MIM stacks. Without carefully engineered tapers, the excited modes suffer loss due to power diverted into non-targeted modes, including radiative modes. Poorly designed tapers also contribute to low extinction ratios as unmodulated modes propagate forward and interfere with the modes emerging from the taper section. Thus, optimal performance critically depends on the successful transformation of the $TM_0$ mode of the input Si waveguide to the MIM stacks via the input taper, and effectively returning the modulated mode back to the $TM_0$ mode of the output Si waveguide via the output taper.

The dimensions and geometric configuration of the taper, specifically its length and angle, are critical parameters that determine how effectively the mode is transformed (minimal losses and undesired mode coupling). We examined the performance of various taper designs by conducting 3D simulations of back-to-back tapers following the sketch of Fig. 1(a) while varying the taper length at flat-band voltage such that the carrier density within the ITO remains unperturbed. In this evaluation, the gap of the input taper changed linearly from 520 nm at its input to $g_{min} = 340$ nm in the MIM section. For the output taper, the gap increased from 340 nm in the MIM section to 520 nm at its output.

Fig. 3 shows the transmittance of the $TM_0$ mode of the Si waveguide through back-to-back tapers, along with its transmittance into all forward propagating modes, and its reflectance into all backward propagating modes, *vs*. the length of each back-to-back taper, $L_T$. The output powers were computed at the output port and normalized to the forward propagating power carried by the $TM_0$ mode of the Si waveguide at the input, to yield transmittances. The total backward propagating power was computed at the input of the structure and similarly normalized to give the reflectance. As shown, there is negligible back reflection for $L_T > 2.5$ µm, whereas for $L_T \leq 2.5$ µm, back reflection is observed, reaching 0.06 for $L_T = 1$ µm. Importantly, at the taper length of 3.1 µm, the $TM_0$-to-$TM_0$ transmittance reaches its maximum value of 0.229, corresponding to an insertion loss of 6.4 dB. Coupling of the $TM_0$ mode to other forward-propagating modes is minimized at this length, as observed by the minimal difference between the blue and red curves.

The electric field distribution at the input, middle, and output of the modulator, corresponding to transmission peaks in Fig. 3 (for taper lengths of $L_T = 1.75, 3.1$ and $4.5$ µm), is shown in Fig. S1. As shown in Fig. 3 and Fig. S1, a small taper size can cause part of the $TM_0$ mode to couple into undesirable weakly-bound modes during propagation along the modulator. To mitigate this issue, we introduce four S-bend silicon (Si) waveguides at the end of the modulator, effectively removing the weakly-bound forward-propagating modes by inducing radiation loss, as discussed in Subsection S.2, Fig. S2 of the Supplemental Material.

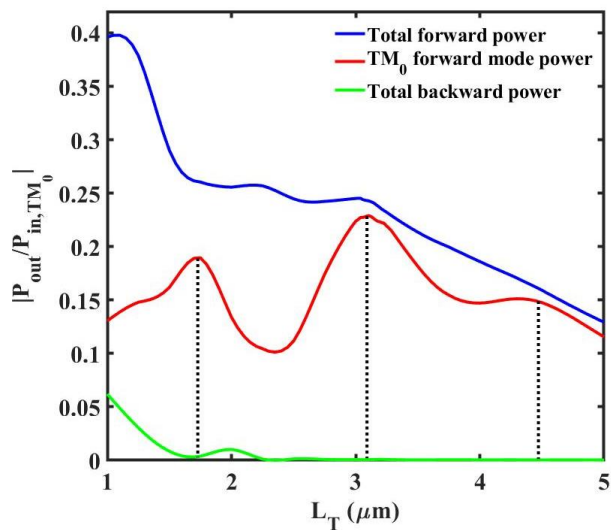

**Fig. 3.** Normalized total forward propagating power at the output (transmittance), normalized $TM_0$ forward propagating mode power at the output (transmittance), and normalized total backward propagating power at the input (reflectance) computed as a function of taper length, $L_T$, for back-to-back tapers (Fig. 1(a)) biased at the flat band voltage. At the taper length of $L_T = 3.1$ µm, minimal $TM_0$-to-$TM_0$ insertion loss is identified.

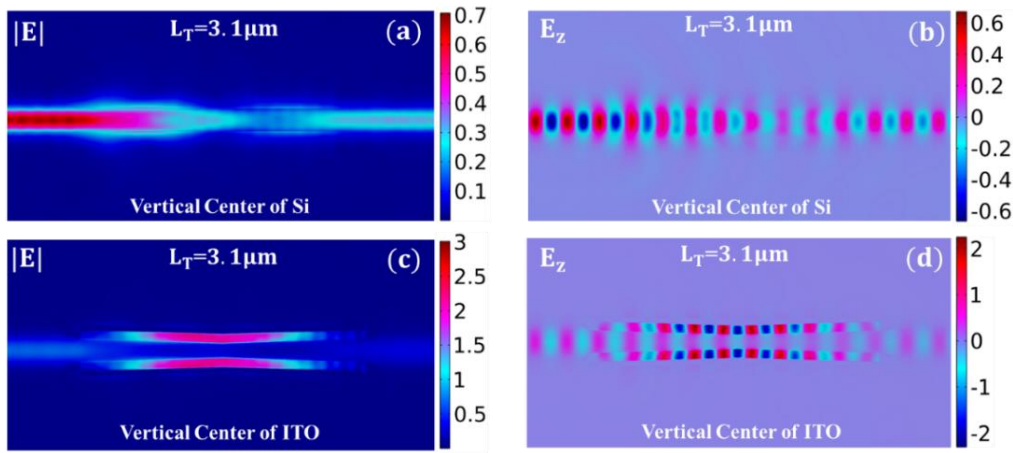

**Fig. 4.** Electric field magnitude (left) and its main component (z-component, right) for the taper length of $L_T = 3.1$ µm. (a) and (b) show the magnitude and z component of the electric field, respectively, at the vertical center of the silicon waveguide, while (c) and (d) show them at the vertical center of the ITO.

Fig. 4 plots the electric field distribution at two longitudinal planes for the taper length of $L_T = 3.1$ μm biased at the flat band voltage, providing detailed insight into mode confinement and transformation along the modulator structure. Figs. 4(a) and 4(b) show the magnitude and the z-component of the electric field along the vertical center of the Si waveguide, whereas Figs. 4(c) and 4(d) plot these same quantities along the vertical center of the ITO. These plots clearly illustrate effective $TM_0$ mode transformation from the input Si waveguide to the MIM section, localization and intensification of the field in the MIM section, and its subsequent transformation to the $TM_0$ mode of the output Si waveguide. The z-component of the electric field clearly shows that the $TM_0$ mode of the input Si waveguide is transformed to the symmetric TM mode of the MIM section.

The analysis of modal evolution and transformation for various taper lengths provides an understanding of the intrinsic performance of the modulator under flat band conditions. The next subsections examine the effects induced by applying a voltage on mode transformation, field distributions and the performance of the device by perturbing the carrier density within the ITO. This analysis provides insight into modulator dynamics and performance.

*4.3 Bias Dependent Carrier Density of ITO*

To study the effects of the applied voltage on the optical response of the device, we explore the performance of the MOS cross-section comprising Al, $Al_2O_3$ and ITO, as shown in Fig. 5(a), which is embedded within the MIM stacks of the modulator. When voltage is applied across this structure, the carrier density in the ITO layer is altered, transitioning from depletion, through flat band, to accumulation, based on the polarity and strength of the applied voltage.

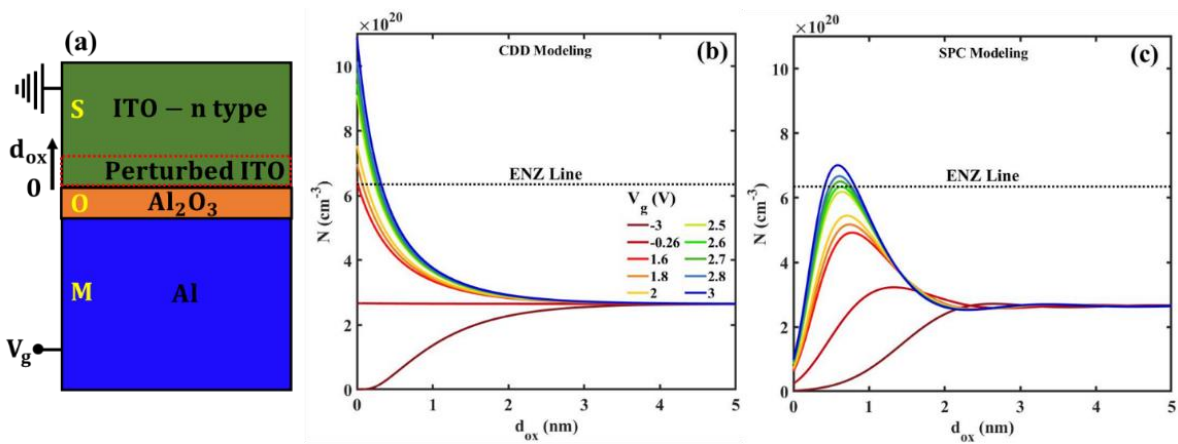

**Fig. 5.** Influence of the voltage applied to the MOS structure, $V_g$, on the electron density profile, N, within ITO. (a) MOS structure schematic. (b) CDD modeling results showing voltage-induced variations in electron density. (c) SPC modeling showing pronounced effects due to quantization, leading to two ENZ crossings and oscillations in electron density.

Fig. 5(b) presents the electron density distribution within the perturbed region of ITO at various gate voltages, predicted by the CDD model. At a gate voltage of -3 V, the model yields a depletion region of significantly reduced carrier density, which extends from the oxide-ITO interface to about 2 nm

10 | P a g e

into the ITO. As the voltage increases to -0.26 V, the system reaches the classical flat band condition, characterized by neither accumulation nor depletion of carriers. Upon increasing the voltage beyond -0.26 V, the structure transitions to accumulation. Electron accumulation, most prominent at the oxide-ITO interface, follows an exponential decay into about 2 nm of the ITO. At a gate voltage of 1.6 V, the carrier density at this interface reaches the epsilon-near-zero (ENZ) threshold indicated by the black dotted line. As the voltage continues to rise, the carrier density surpasses the ENZ line at the interface, indicating a shift in material properties from dielectric to metallic. This transition is critical as it fundamentally alters the optical properties of the accumulation region (dielectric to metallic) and is key to the performance of the modulator.

Fig. 5(c) illustrates the variations in electron density within the ITO as determined by the SPC model for a range of applied gate voltages. Similarly to the CDD model, at $V_g = -3$ V, the SPC model predicts a depletion region at the oxide-ITO interface. However, the SPC model predicts a significantly lower electron density at this interface for all applied voltages, leading to a pronounced bump in density further into the ITO. This bump becomes increasingly pronounced as the applied voltage increases. As shown in Fig. 5(c), the bump in electron density reaches the ENZ line at 2.6 V and crosses it at two distinct locations within the ITO as the voltage increases. Additionally, the distribution of the electron density exhibits Friedel-like oscillations at higher voltages due to electron diffraction and interference [39]. As further discussed below, this complex behavior - a significant reduction in carrier density at the oxide-ITO interface, dual ENZ crossings, and oscillatory effects - predict significant differences on the modulator's optical response relative to the CDD model.

We applied the voltage-induced perturbed carrier density to the Drude model, Eq. (4), to obtain the distribution of the real and imaginary parts of permittivity within the perturbed region of the ITO. For voltages greater than the ENZ voltages ($V_{ENZ}^{CDD} = 1.6$ V for the CDD model and $V_{ENZ}^{SPC} = 2.6$ V for the SPC model) the real part of permittivity turns negative, revealing the formation of a metallic layer inside the ITO. The real and imaginary parts of permittivity at different gate voltages are given in Fig. S3 of the Supplemental Material.

To explore the impact of these permittivity changes on the effective refractive index ($n_{eff}$), we performed 2D modal analyses on the device cross-section (Fig. 1b) for different gate voltages and gap widths, g. Fig. S4 (Supplemental Material) shows how $n_{eff}$ varies as the gap decreases from 520 nm to 340 nm across different gate voltages. Fig. 6 highlights the behavior of $n_{eff}$ at a fixed gap width of 340 nm, for both the CDD and SPC models. Both models show a reduction in Re($n_{eff}$) as the voltage approaches their respective ENZ voltages (1.6 V for CDD, 2.6 V for SPC), followed by an increase beyond these voltages. Specifically, the change in Re($n_{eff}$) predicted by the SPC model is more than double that predicted by the CDD model. For Im($n_{eff}$), both models behave similarly up to 1 V. Beyond this, Im($n_{eff}$) rises more sharply in the SPC model, peaking at 2.7 V, while the CDD model shows saturation after 1.8 V. The SPC model predicts a ~5× increase in Im($n_{eff}$) compared to depletion ($V_g = -3$ V), while the CDD model shows a ~2× increase.

These discrepancies between the CDD and SPC models arise from differences in the electric field behavior within the perturbation region [20]. The CDD model, with a single ENZ point, results in one sharp peak and moderate changes in $n_{eff}$. In contrast, the SPC model, with two ENZ points, produces two distinct peaks and larger changes due to a broader field distribution.

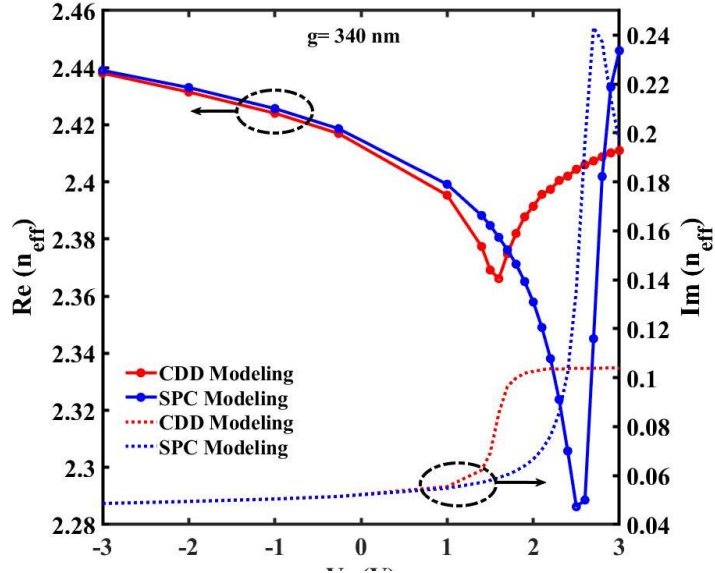

**Fig. 6.** Real part (left axis) and imaginary part (right axis) of $n_{eff}$ of the MIM coupled mode for $g = 340$ nm *vs.* applied gate voltage, $V_g$, with the perturbation in ITO modelled using the CDD and SPC models.

*4.4 Extinction Ratio and Field Distribution at Various Gate Voltages*

The extinction ratio (ER) is a critical performance metric of electro-optic modulators, as it quantifies the contrast between the on and off states of an intensity modulated signal. A high ER indicates effective modulation, with a clear distinction between the transmitted and blocked states. To determine the ER, we performed 3D optical simulations of the full structure, consisting of input and output Si waveguides and back-to-back tapers forming the modulator, similarly to Subsection 4.2 but at different gate voltages. The simulations involved casting the perturbed permittivity of the ITO computed at various voltages (*cf.* Fig. S3) to the 3D optical model of the full structure and obtaining the $TM_0$-to-$TM_0$ transmittance using the Beam Envelope Method (*cf.* Subsection 3.2). The extinction ratio is calculated using the formula:

$$ER = 10\, log\left(\frac{P_{on}}{P_{off}}\right) \tag{11}$$

where $P_{on}$ is the output power carried by the $TM_0$ mode of the Si waveguide at a voltage of -3 V, corresponding to strong depletion inside the perturbed region of ITO (and minimum insertion loss), and $P_{off}$ is the output power carried by the same mode at higher voltages.

Fig. 7 shows the $TM_0$-to-$TM_0$ ER (left axis) and transmittance (right axis) at different gate voltages, computed using the CDD and SPC models, for a modulator of taper length $L_T = 3.1$ µm. The SPC model predicts that the ER increases with increasing voltage, reaching its maximum at 2.8 V (slightly above $V_{ENZ}^{SPC}$), beyond which it decreases. This behavior follows that of $Im(n_{eff})$, as plotted in Fig. 6. The CDD model, alternatively, shows an increase in ER with increasing voltage, reaching a maximum value at around 2 V after which it becomes almost constant, also following the corresponding trend in $Im(n_{eff})$ plotted in Fig. 6. A significant difference between the models is the prediction of a maximum ER of ~25 dB for the SPC model, compared to a maximum of ~7 dB for the CDD model. This difference is due ultimately to the different behavior of the carrier distributions in the perturbed region of ITO of these models (*cf*. Fig 5). This analysis highlights the importance of considering quantization of carrier energy states in the modeling of plasmonic electro-optic modulators based on ENZ materials.

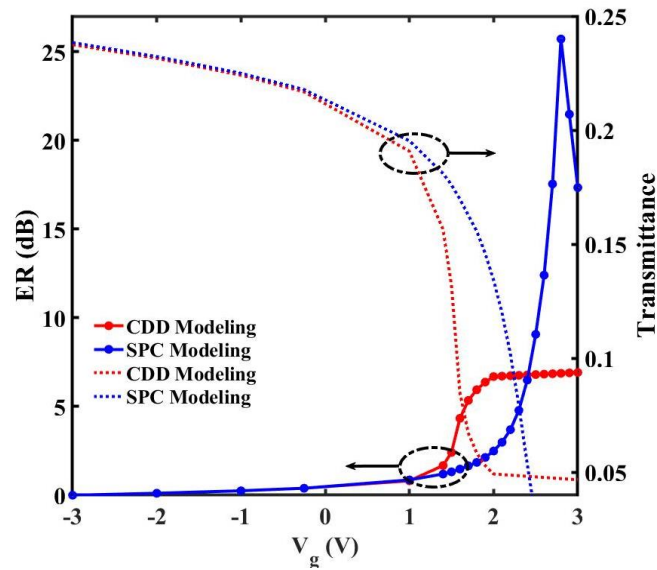

**Fig. 7.** $TM_0$-to-$TM_0$ extinction ratio (left axis) and transmittance (right axis) of the modulator *vs*. gate voltage, computed using the CDD and SPC models. The taper length is set to $L_T = 3.1$ µm.

To further elucidate the effects of the voltage, we examine field distributions along the modulator at three different gate voltages, $V_g$, computed using our 3D model of the structure. The electric fields are normalized at the input and maintain their relative magnitude along the length of the structure. Fig. 8(a) shows the distribution of the electric field magnitude at -3 V, for operation under strong depletion corresponding to the low insertion loss (on) state of the modulator. It is noted that the field distribution is almost identical for the CDD and SPC models, given that the models defer little in depletion. We also plot in Figs. 8(b) and 8(c) the field distributions at a gate voltage inducing strong accumulation, just beyond the ENZ voltages, 2 V for the CDD model and 2.8 V for the SPC model, respectively, where the ER reaches maximizes for each model. These voltages correspond to the high insertion loss (off) state of the modulator. The left, middle and right panels show the distribution of the electric field magnitude along the longitudinal vertical center of the silicon waveguide, along the longitudinal vertical center of the ITO, and over the cross-section at the output, respectively.

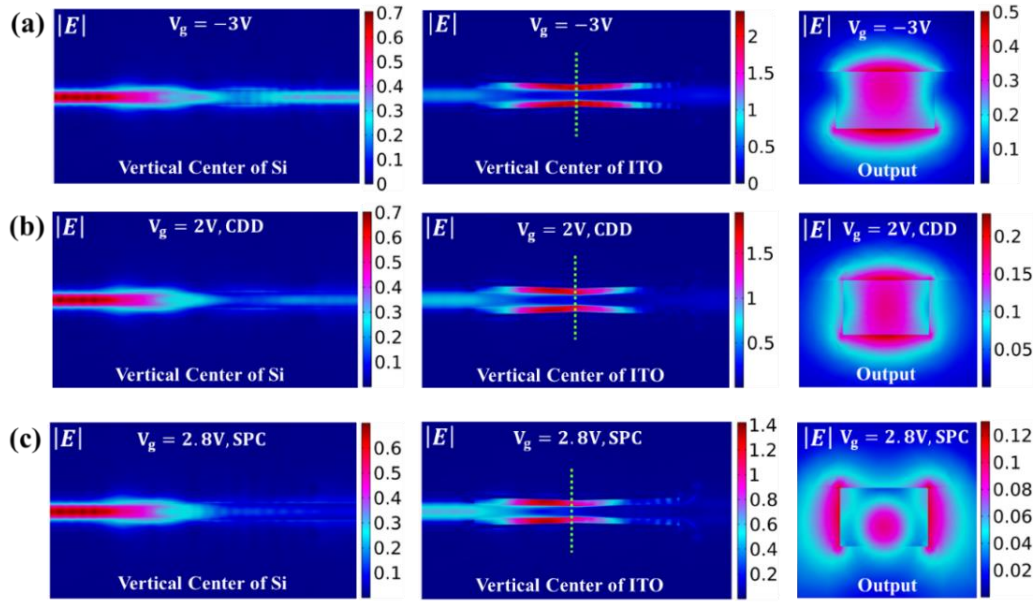

**Fig. 8.** Electric field magnitude distribution of the plasmonic TM mode at the vertical center of Si (left) and ITO (middle), and at the output (right). The distributions are shown at different gate voltages: (a) -3 V (first row) using either model, (b) 2 V (second row) using the CDD model, and (c) 2.8 V (third row) calculated using the SPC model. The taper length is $L_T = 3.1$ μm.

Comparing the field distribution along the Si waveguide under depletion in Fig. 8(a) to those under accumulation in Figs. 8(b) and 8(c) shows the depth of the extinction predicted by the models due to the increased attenuation ($\text{Im}(n_{\text{eff}})$) of the MIM section - in both cases, an attenuated field is noted in the output Si waveguide. The extinction is far more pronounced for the SPC model for the reasons discussed earlier with regards to Fig. 7. Also, the maximum field in the MIM section shifts toward the input at these voltages for both models, illustrating how the applied voltage impacts mode transformation and confinement within the modulator.

It is also noted that the output field distribution under accumulation differs from that of the input $TM_0$ mode (*cf.* Fig. S1), especially for the SPC model. This difference occurs because when the $TM_0$ mode of the silicon waveguide enters and progresses through the tapers of the modulator, it excites (slightly) other unwanted modes (*cf.* Fig. 3). Despite carrying a small fraction of the total power, these unwanted modes propagate through the modulator unaffected by the applied voltage. Consequently, the output field comprises the $TM_0$ mode, which was highly attenuated by the applied voltage, and the residual unwanted (and unmodulated) modes. The unwanted modes are weakly bound and can be eliminated from the output by adding S-bend Si waveguides to the end of the modulator, as discussed relative to the flat band voltage case (Subsection 4.2) and shown in Fig. S2 of the supplementary materials. It is important to note that while these unwanted modes are also present in the CDD model calculations, their contribution to the overall field distribution is relatively smaller than that of the $TM_0$ mode because the former is less attenuated.

*4.5 Bandwidth Analysis*

The electrical bandwidth of a modulator is another critically important parameter as it determines the bit rate at which data can be impressed into the optical carrier. The modulator being very short (~ 6 µm) is considered as a lumped element, so its 3-dB bandwidth is simply that of MOS capacitor driven into accumulation embedded into parasitic and load resistances:

$$BW = \frac{1}{2\pi\left(R+\frac{R_S R_L}{R_S+R_L}\right)C} \quad (12)$$

where R is the parasitic series resistance of the modulator, $R_S$ and $R_L$ are the source and load resistances set to 50 Ω, and C is the MOS capacitance of the modulator under accumulation.

The parasitic resistance was evaluated using the electrostatic model in COMSOL 6.1. In our calculations, the model was simplified by assuming a 2D cross-section of modulator, excluding the oxide layer, which yielded a parasitic resistance corresponding to 3.35 Ω µm. This result aligns closely with the resistance of 3.45 Ω calculated using the formula R = ρL/A for a 1 µm long section, thus validating our simulation methodology.

To obtain the capacitance, we calculated the charge per unit length *vs.* voltage using the CDD and SPC models, as plotted in Fig. S5 (Supplemental Material), yielding values from both approaches that are very close to each other. The capacitance per unit length, derived from the slope of these graphs, is approximately 8.1 fF/µm. This estimate is close to the value 9.2 fF computed from $(\varepsilon_0 \varepsilon_{r,ox} A)/h_{ox}$ for a 1 µm long section, where $\varepsilon_0$ is the permittivity of free space, $\varepsilon_{r,ox}$ is the relative permittivity of alumina (set to 9.3), $h_{ox}$ is the thickness of alumina (taken to be 3.6 nm), and A is the area of both MIM structures. For a modulator of taper length $L_T = 3.1$ µm, the 3-dB bandwidth is thus 124 GHz as computed using Eq. (12).

*4.6 Optimization of Device Performance*

Since the ER obtained with the design above was quite large at more than 25 dB, the modulator performance can be optimized by sacrificing ER to boost the bandwidth and reduce the insertion loss (IL). To achieve this, we maintained the angle of the taper to the same value as above but shrunk the modulator length by increasing the gap g in the MIM section. This reduction in length decreases the area of the MIM section, which in turn reduces the capacitance and increases the bandwidth. Additionally, a shorter length reduces the insertion loss. However, these improvements come at the cost of a decreased extinction ratio, as we present below.

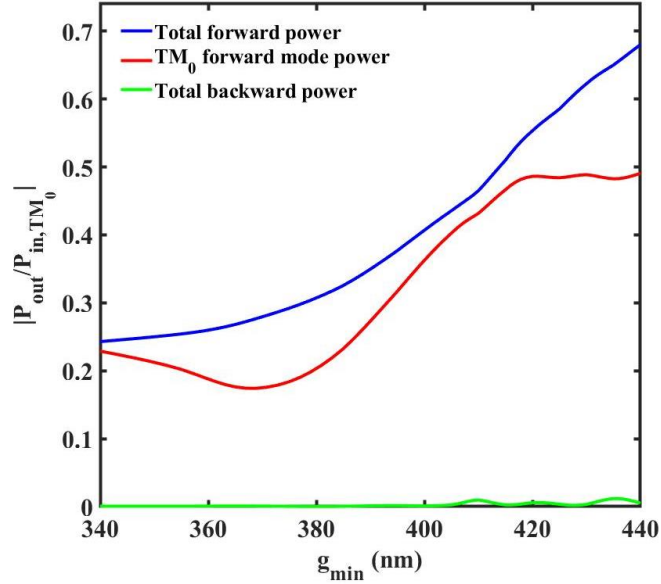

**Fig 9.** Normalized total forward propagating power at the output (transmittance), normalized $TM_0$ forward propagating mode power at the output (transmittance), and normalized total backward propagating power at the input (reflectance) computed as a function of gap g between MIM sections, for back-to-back tapers (Fig. 1(a)) biased at the flat band voltage.

Fig. 9 shows the transmittance of the $TM_0$ mode of the Si waveguide through back-to-back tapers, along with its transmittance into all forward propagating modes, and its reflectance into all backward propagating modes, *vs.* the minimum gap size in the MIM region, $g_{min}$. The tapers are all at the same angle, but of different lengths depending on $g_{min}$. The output powers were computed at the output port and normalized to the forward propagating power carried by the $TM_0$ mode of the Si waveguide at the input, to yield transmittances. The total backward propagating power was computed at the input of the structure and similarly normalized to give the reflectance. These results were obtained at the flat band voltage and computed using our 3D optical model of the structure. As shown in this figure, the back-reflected power is negligible for all MIM gaps. The difference between the total forward propagating power and the $TM_0$ forward propagating power at the output indicates the fractional coupling to unwanted modes, which is minimized at $g_{min} = 340$ nm and for g in the range of 400 to 415 nm.

We conducted 3D optical simulations at different gate voltages for MIM gaps of $g_{min} = 400$ and 415 nm, and compared the results to those obtained previously for $g_{min} = 340$ nm. The results are summarized in Table 1, including the gate voltages at which the ER is maximum as computed using the CDD and SPC models. As mentioned earlier, increasing the gap reduces the length of the taper and the area of the MIM section. This reduction results in increased bandwidth and decreased insertion loss, although it also leads to a reduced ER, as noted from the data in the Table. Using shorter modulators leads to $TM_0$-to-$TM_0$ IL in the range of 3-5 dB, ER of 5-10 dB (SPC), and an electrical BW of 180-210 GHz. Additionally, as the gap increases, the drive voltage required to achieve the maximum ER in the CDD and SPC models decreases, reaching their minimum value, which is equal to the ENZ voltage for each model. The field distributions for MIM gaps of $g_{min} = 400$ and 415 nm, computed at the flat band voltage and at the voltages producing the largest extinction ratio, are shown in Figs. S6 and S7 of the supplementary materials. Using shorter modulators with gaps in the range of 400 - 415 nm achieves a good balance between the device performance characteristics, in the form of increased bandwidth and decreased insertion loss with a good extinction ratio.

Table 1. Optimized ER, BW, and IL for different gaps obtained via the CDD and SPC modeling.

| Gap (nm) | $L_T$ (µm) | $IL_{total}$ (dB) | BW (GHz) | $V_g$ (V) - CDD | $ER_{max}$ (dB) - CDD | $V_g$ (V) - SPC | $ER_{max}$ (dB) - SPC |
|---|---|---|---|---|---|---|---|
| 340 | 3.10 | 6.4 | 123.9 | 2.0 | 6.8 | 2.8 | 25.7 |
| 400 | 2.05 | 4.6 | 185.4 | 1.9 | 3.9 | 2.6 | 9.8 |
| 415 | 1.80 | 3.3 | 210.3 | 1.6 | 3.1 | 2.6 | 5.4 |

## 5. Conclusions

In this study, we introduce a design of a high-speed plasmonic electro-optic modulator with a high extinction ratio, operating at a free-space operating wavelength of 1550 nm. These modulators address significant challenges in existing designs through a combination of innovative structural and material modifications. The modulator, comprised of a pair of coupled Metal-Insulator-Metal (MIM) waveguides integrated on a planarized Si waveguide, benefits from reduced structural complexity, which enhances fabrication feasibility and device performance by minimizing parasitic effects. The coupled MIM structures act as tapers, adiabatically transforming the $TM_0$ mode of the input Si waveguide to the symmetric plasmonic TM mode in the MIM section, and re-transforming it to the $TM_0$ mode of the output Si waveguide. By applying a voltage, we induce epsilon-near-zero (ENZ) in the perturbed region of ITO, of $Al - Al_2O_3 - ITO$ MOS structures embedded in the MIM stacks, by driving the latter into strong accumulation.

We used the Classical Drift-Diffusion (CDD) and Schrödinger-Poisson Coupling (SPC) methods models to obtain voltage-dependent perturbed carrier density inside the ITO and determine the electro-optical response of the modulator. These models predict similar results in the depletion region but differ significantly in the accumulation regime. Specifically, in strong accumulation, the SPC model predicts two ENZ points for the $Re(\varepsilon)$ of the perturbed region of the ITO, while the CDD model predicts a single ENZ point, resulting in higher changes in the optical response of the device predicted by the SPC model. For example, $\Delta Re(n_{eff})$ and $\Delta Im(n_{eff})$ predicted by the SPC model are -0.155 and 0.19 for an applied voltage attaining ENZ, which is ~2× and ~4× the corresponding values predicted by the CDD model. These differences highlight the importance of effects due to quantization on optical response of the modulator that are contained within the SPC but not the CDD, which need to be considered for accurately computing the perturbed carrier density within the ITO.

One optimal linear taper design of length $L_T = 3.1$ µm yields a modulator that achieves a $3 - $ dB bandwidth of 125 GHz, an insertion loss of 6 dB, and an extinction ratio of 26 dB as predicted by the SPC model. The trade-off between bandwidth and insertion loss *vs*. extinction ratio was also analyzed in detail, resulting in another optimal design that yields a $3 - $ dB bandwidth of 210 GHz and an insertion loss of 3 dB, for a taper length of $L_T = 1.8$ µm, but at a lower ER of 5 dB, as predicted by the SPC model. Our results offer insights for further optimization, demonstrating the balance between high-speed operation, low insertion loss, and modulation depth. The modulator designs are suitable for high-speed optical interconnects, offering scalable solutions for integrated photonics and optical communications.

**Funding.** Financial support provided by Huawei Canada is gratefully acknowledged.

**Acknowledgments.** We would like to acknowledge CMC Microsystems and Canada's National Design Network for access to COMSOL Multiphysics 6.1 licenses.

**Disclosures.** The authors declare no conflicts of interest.

**Data availability.** All data underlying the results or needed to evaluate the conclusions of the paper are present in the paper and/or Supplement 1.

**Supplemental document.** See Supplement 1 for supporting content.

# Plasmonic Electro-Optic Modulators based on Epsilon-Near-Zero Materials: Comparing the Classical Drift-Diffusion and Schrödinger-Poisson Coupling Models

**MASOUD SHABANINEZHAD,[1,2,3] HAMID MEHRVAR,[4] ERIC BERNIER,[4] LORA RAMUNNO,[2,3] AND PIERRE BERINI[1,2,3]**

[1]School of Electrical Eng. and Comp. Sci., University of Ottawa, Ottawa, ON K1N 6N5, Canada
[2]Department of Physics, University of Ottawa, Ottawa, ON K1N 6N5, Canada
[3]NEXQT Institute, 25 Templeton St., Ottawa ON, K1N 6N5, Canada
[4]Huawei Technologies Canada, Canada, Kanata, ON K2K 3J1, Canada

**S. 1. Cross-Sectional Electric Field Distribution at Transmission Peaks**

Fig. S1 shows the top view of the modulator and the cross-sectional distribution of the electric field plotted at the input, the middle of the MIM section, and the output, for three taper designs of length $L_T = 1.75$, 3.1 and 4.5 μm where peaks appear in the $TM_0$-to-$TM_0$ transmittance of Fig. 3.

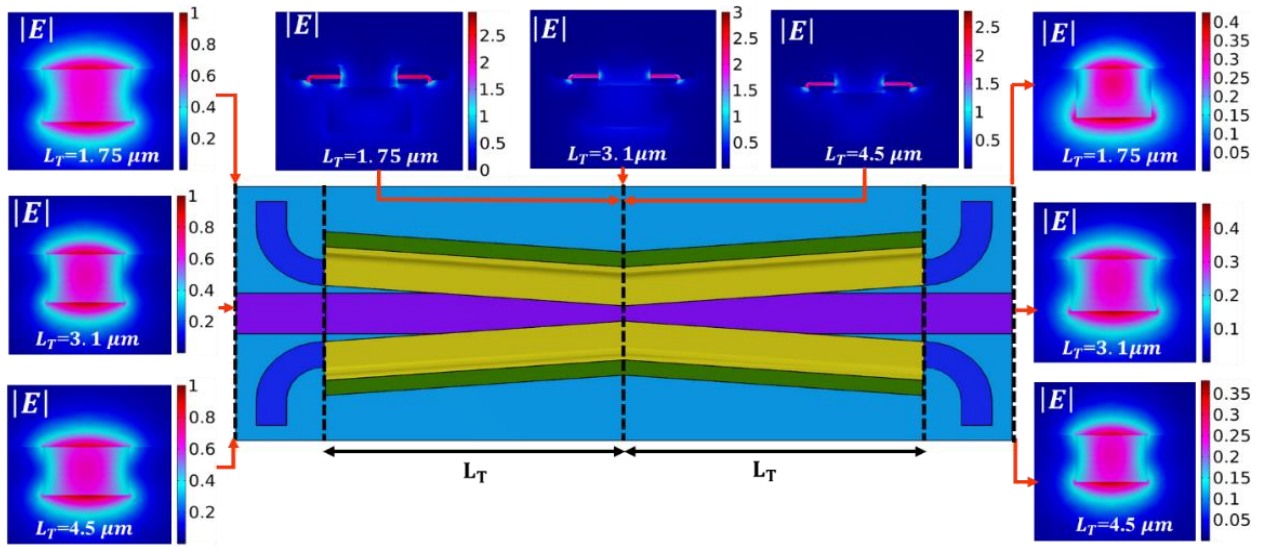

**Fig. S1.** Top view of the modulator and cross-sectional electric field distributions at the input, middle of the MIM section, and output of the structure for three taper lengths: $L_T = 1.75$, 3.1 and 4.5 μm. The field distributions at the output for $L_T = 3.1$ and 4.5 μm closely resemble that at the input, indicating minimal coupling into other forward propagating modes.

The electric fields are normalized at the input and maintain their relative magnitude along the length of the structure. In these plots, the $TM_0$ mode is launched in the Si waveguide toward the input, transfers adiabatically to the MIM section via the input taper, then returns to the Si waveguide through the output taper. The field distribution in the middle of the MIM section shows that the field is well-confined and enhanced therein relative to the input, indicating efficient mode transformation from the Si waveguide to the MIM section. For the taper lengths of 3.1 and 4.5 µm, the field distribution at the output closely resembles that at the input. However, for the taper length of 1.75 µm, the field distribution at the output differs slightly from that at the input, indicating not only coupling to the output $TM_0$ mode but also to other weakly-bound forward propagating modes. This observation is consistent with the transmittance computations of Fig. 3.

**S. 2 S-Bend Silicon Waveguides**

For a given gap, when the size of the taper is small, some part of the $TM_0$ mode will couple to undesired modes while propagating along the modulator. To remove these undesirable modes that are weakly bounded, we add four S-bend Si waveguides to the end of the modulator.

Fig. S2 shows the top view of the modulator, including the S-bend silicon waveguides, and the electric field distribution of the $TM_0$ mode of the Si waveguide at the input port (left panel), output port (middle panel), and end of the S-bend Si waveguides (right panel). The simulation was done for the flat band voltage case, and the length of the taper, and radius of the S-bend waveguides were taken as $L_T = 1$ µm and $r_S = 2$ µm, respectively. As shown, the field distribution at the output port (middle panel) significantly differs from the input one; however, at the end of the S-bend waveguides, it closely resembles the input one, demonstrating the effectiveness of this approach for stripping weakly bound higher order modes.

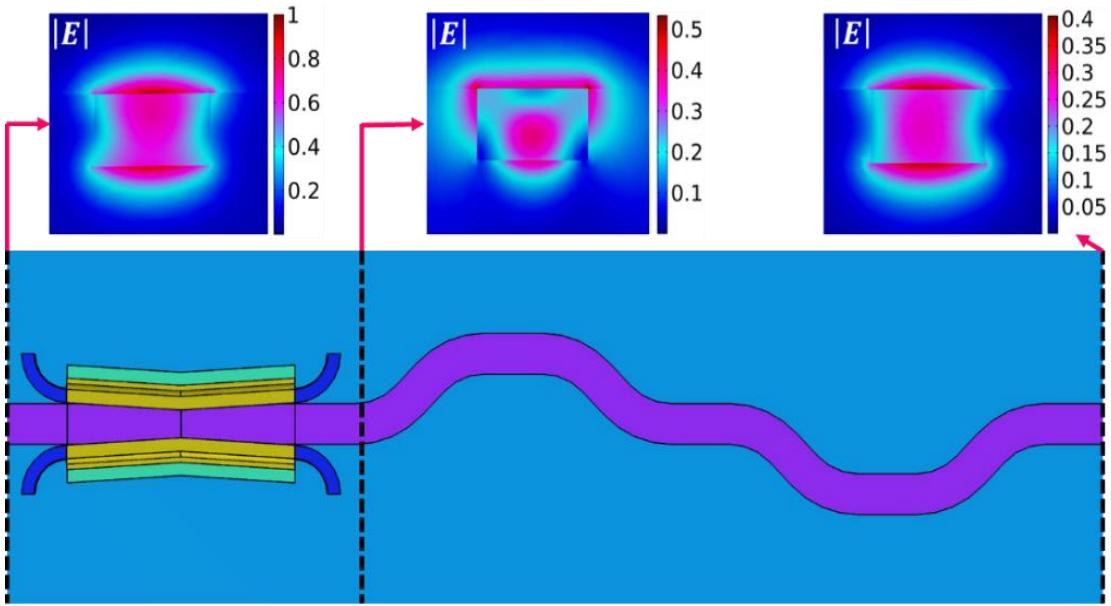

**Fig. S2.** Electric field distribution of the $TM_0$ mode of the Si waveguide for a modulator with $L_T = 1$µm at the flat band voltage. The left, middle and right panels show the magnitude of the electric field distribution at the input port, output port, and after the S-bend silicon waveguides, respectively.

## S. 3 Bias Dependent Permittivity of ITO

Figs. S3(a) and S3(b) show the real part of the permittivity within the perturbed region of ITO at different gate voltages obtained using the CDD and SPC models, respectively. The bulk real part of the permittivity of ITO is 2.4, reaching about 4 in the depletion region. As the voltage increases, the real part of the permittivity in the perturbed region decreases, reaching zero at the ENZ voltages: $V_{ENZ}^{CDD} = 1.6$ V for the CDD model and $V_{ENZ}^{SPC} = 2.6$ V for the SPC model. With further voltage increase, the real part of permittivity becomes negative, indicating the creation of a metallic layer inside the ITO. The red semi-transparent rectangle in Figs. S3(a) and S3(b) highlights the metallic region created inside the ITO at $V_g = 3$ V, of thickness 0.5 nm for the CDD model and 0.7 nm for the SPC model. Notably, the CDD model predicts that the metallic layer starts from the oxide-ITO interface and extends into the ITO for gate voltages greater than ENZ voltage, whereas the SPC model predicts that this layer is located away from the oxide-ITO interface.

Figs. S3(c) and S3(d) give the imaginary part of the permittivity in the perturbed region of the ITO, computed using the CDD and SPC models, respectively. Both figures reveal a trend that closely aligns with the carrier density distribution, demonstrating how increased voltage leads to a rise in the imaginary part of the permittivity, signifying enhanced metallic properties within the perturbed region.

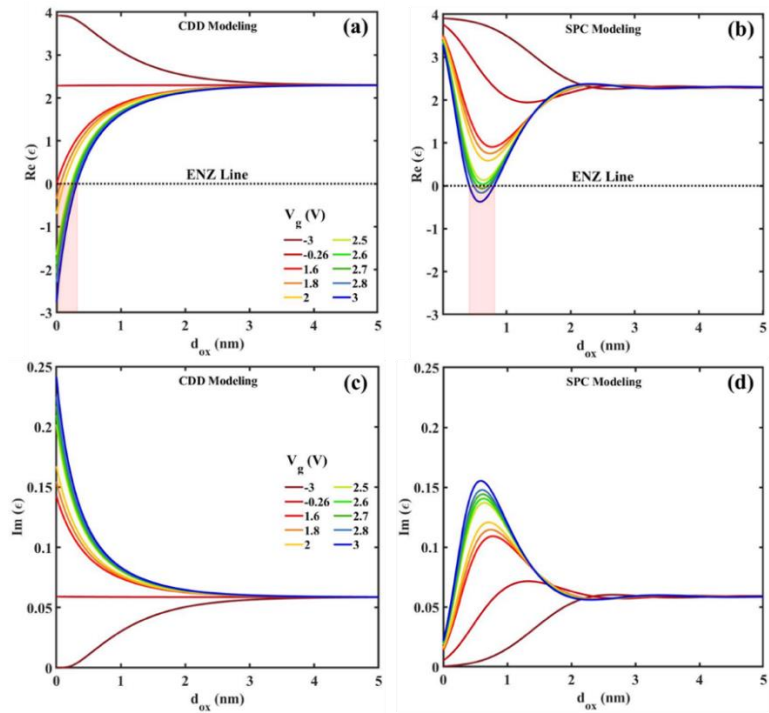

**Fig. S3.** Voltage-induced spatially-dependent permittivity within the ITO. (a) and (b) Voltage-dependent real part of permittivity computed using the CDD and SPC models, highlighting the transition to metallic properties at ENZ voltages (red shading). (c) and (d) Voltage-dependent imaginary part of permittivity computed via the CDD and SPC models.

## S.4 Modal Transformation Under Applied Bias

Fig. S4(a) illustrates the variations in the real part of the effective refractive index ($Re(n_{eff})$) of the symmetric MIM mode as the gap decreases from g = 520 to 340 nm, using the CDD model at different gate voltages. Notably, $Re(n_{eff})$ generally increases as the gap narrows for all gate voltages. For narrow gaps, say g < 400 nm, $Re(n_{eff})$ decreases as the gate voltage increases from -3 to 1.6, then reverses beyond this range and begins to saturate for $V_g > 1.8$ V. At $V_g = V_{ENZ}^{CDD} = 1.6$ V, the carrier density in the perturbed region of ITO reaches the ENZ line, causing the real part of the permittivity to drop to zero and $Re(n_{eff})$ to reach a minimum near this voltage. Beyond 1.6 V, the formation of a metallic layer in the perturbed region expels field from this region, and consequently $Re(n_{eff})$ begins to rise again. Beyond 1.8 V, $Re(n_{eff})$ approaches saturation, showing negligible further increases. Conversely, $Im(n_{eff})$ simply increases with voltage, reaching saturation at $V_g \sim V_{ENZ}^{CDD} = 1.6$ V for narrow gaps, as observed in Fig. S4(c).

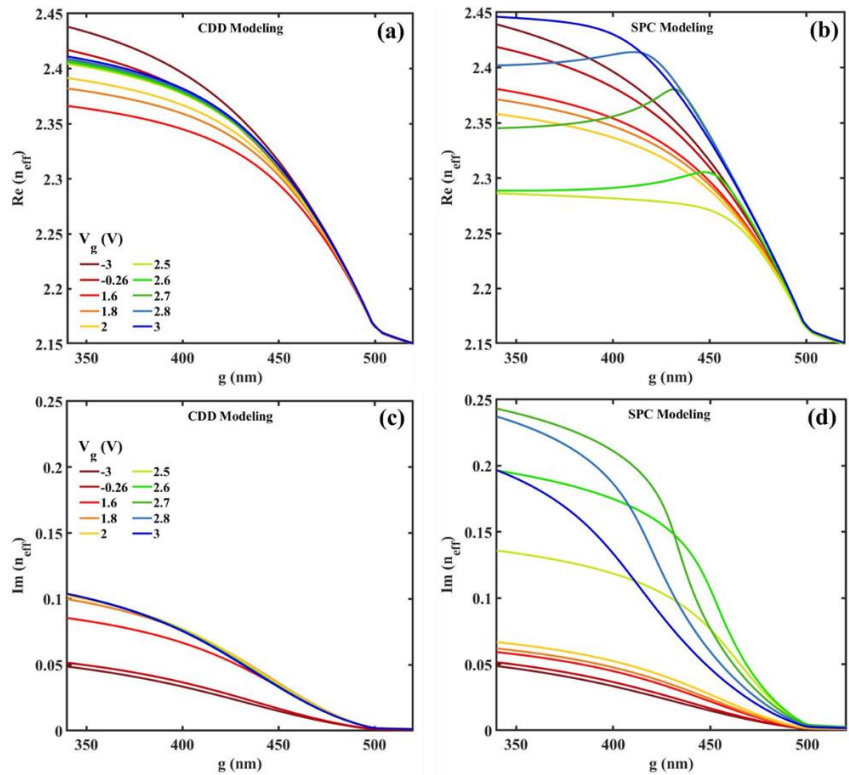

**Fig. S4.** Evolution of the fundamental TM mode in the Si waveguide into the MIM coupled mode with gap width, g, at various applied gate voltages, $V_g$, (legend) with the perturbation in ITO modelled using the CDD and SPC models. (a) $Re(n_{eff})$ and (c) Im ($n_{eff}$) obtained from modal analysis using the CDD model. (b) Re ($n_{eff}$) and (d) Im ($n_{eff}$) obtained

The SPC model predicts a markedly different behavior for $Re(n_{eff})$ at the same voltages as the gap decreases from g = 520 to 340 nm, as observed by comparing Fig. S4(b) to Fig. S4(a). For narrow gaps, say g < 400 nm, there is a consistent decrease in $Re(n_{eff})$ as the applied voltage increases from -3 V to 2.5 V. At $V_g = V_{ENZ}^{SPC} = 2.6$ V the perturbed carrier density within the ITO reaches the ENZ point and $Re(n_{eff})$ reaches a minimum. Beyond 2.6 V, $Re(n_{eff})$ increases consistently up to 3 V. This behavior is qualitatively similar to that predicted by the CDD model, except that the range over which $Re(n_{eff})$ varies with voltage is larger as predicted by the SPC model.

However, a qualitative difference between the models is observed as the gap is reduced from $g = 520$ nm for gate voltages greater than the ENZ voltage. In this voltage range, $Re(n_{eff})$ initially increases as g decreases, similarly to the CDD model, but reaches a maximum at $g \sim 450$ nm, where the modal character becomes primarily MIM-like. Beyond this gap width, $Re(n_{eff})$ drops slightly as the gap decreases, as opposed to the behavior observed with the CDD model. The differences in these trends are explained by details in the mode fields within the MIM structure, as discussed below.

$Im(n_{eff})$ obtained via the SPC model exhibits a consistent rise as the gap decreases from $g = 520$ nm but the rate of increase changes below ~450 nm as a function of applied voltage, as observed in Fig. S4(d). For narrow gaps, say $g < 400$ nm, $Im(n_{eff})$ increases significantly as $V_g$ approaches the ENZ voltage, reaching a maximum near the latter and decreasing for gate voltages greater than the ENZ voltage. The magnitude of the increase in $Im(n_{eff})$ is significantly larger than that predicted using the CDD model as can be appreciated by directly comparing Figs. S4(c) to S4(d).

## S. 5 Charge per unit length distribution

The charge per unit length as a function of gate voltage was calculated using both the CDD and SPC models (Fig. S5). As shown, the overall charge behavior is nearly identical for both models, confirming the consistency of their predictions for the capacitance of the modulator.

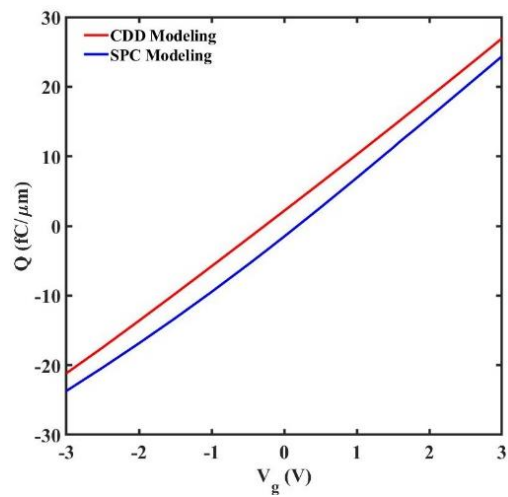

**Fig. S5.** Charge per unit length versus gate voltage for the modulator, comparing results from the CDD and SPC models.

## S. 6 Voltage-Dependent Field Distributions Through the Modulator for Different Modulator Gaps

Fig. S6 presents the field distributions at different gate voltages, normalized by dividing by the maximum field value computed at the input port, for a taper length of $L_T = 2.05$ μm corresponding to a minimum gap of $g_{min} = 400$ nm between the MIM sections. Each row shows the magnitude of the electric field distribution at three different planes: the vertical center of the Si, the vertical center of the ITO, and the output, from left to right, respectively. Fig. S6(a) shows the field distribution at $V_g = -3$ V, which corresponds to strong depletion. The CDD and SPC models predict similar results. Fig. S6(b) displays the field distribution at the same planes at 1.9 V using the CDD model, where the ER reaches its maximum value of 3.9 dB. Fig. S6(c) illustrates the field distribution at 2.6 V obtained using the SPC model, which is the voltage producing the maximum ER of 9.8 dB. The results indicate that application of the voltage attenuates the MIM mode in the modulator, as shown by the reduction in the magnitude of the field distributions. The induced attenuation is more pronounced using the SPC model than the CDD model.

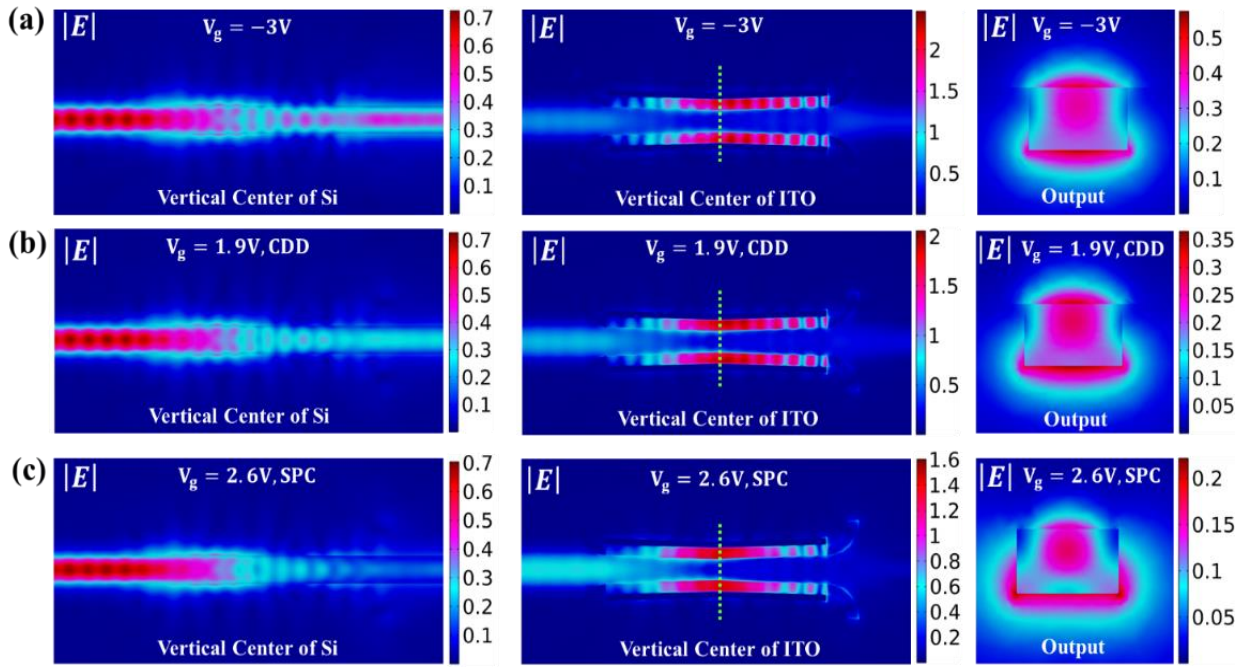

**Fig. S6.** Normalized field distributions at various gate voltages for a modulator of $L_T = 2.05$ μm, corresponding to a minimum gap of $g_{min} = 400$ nm in the MIM section. Each row shows the field at the vertical center of Si and ITO, and at the output, respectively. The electric field distribution are shown at: (a) $V_g = -3$ V (depletion), (b) $V_g = 1.9$ V (CDD), and (c) $V_g = 2.6$ V (SPC).

Fig. S7 presents similar results to Fig. S6, except for $L_T = 1.80$ μm corresponding to a minimum gap of $g_{min} = 415$ nm, and for different voltages. The field distributions are shown for the same planes as in Fig. S6. The required voltage to reach the maximum ER value predicted by the CDD model further reduces to $V_g = 1.6V$, which is the ENZ voltage predicted by this model. By comparing Figs. S7 with S6, it is evident that reducing the length of the taper decreases off state attenuation resulting in a reduced ER, as given in Table 1.

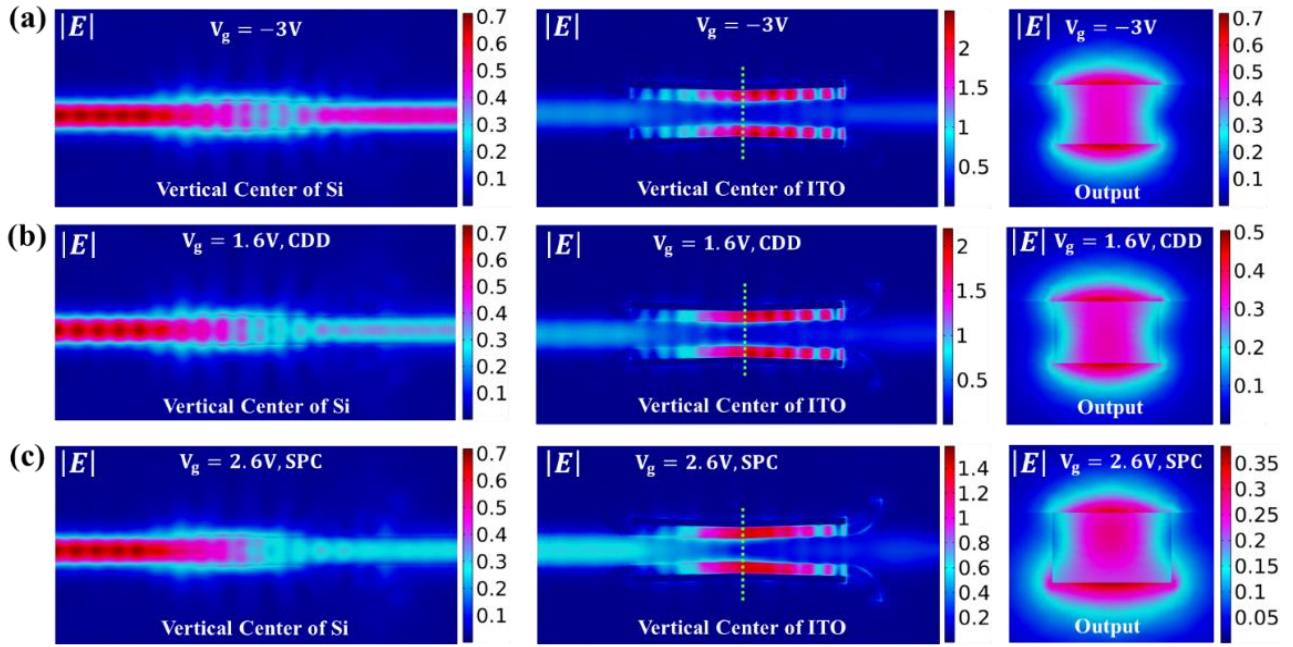

**Fig. S7.** Normalized field distributions at various gate voltages for a modulator of $L_T = 1.80$ μm, corresponding to a minimum gap of $g_{min} = 415$ nm in the MIM section. Each row shows the field at the vertical center of Si and ITO, and at the output, respectively. The electric field distribution are shown at: (a) $V_g = -3$ V (depletion), (b) $V_g = 1.6$ V (CDD), and (c) $V_g = 2.6$ V (SPC), respectively.